\documentclass[reprint,superscriptaddress,pra,nofootinbib,preprintnumbers]{revtex4-2}

\usepackage{graphicx}    
\usepackage{microtype}   
\usepackage{amsmath}     
  \usepackage{amsthm}    
  \usepackage{amssymb}   
\usepackage{siunitx}     
\usepackage{booktabs}    
\usepackage[colorlinks=true,citecolor=blue]{hyperref}

\usepackage{tikz}
  \usetikzlibrary{quantikz}

\usepackage{adjustbox}    
\usepackage{subcaption}
\usepackage{xcolor}
\usepackage{mathtools} 
\usepackage{bm}
\usepackage[shortlabels]{enumitem}  
\usepackage[ruled,vlined]{algorithm2e}

\providecommand{\uone}{{${\rm U}(1)\,\,$}}
\providecommand{\bMax}{{b_{\rm max}}}

\providecommand{\df}{{\rm d}}

\begin{document}

\title{Efficient Representation for Simulating U(1) Gauge Theories on Digital Quantum Computers at All Values of the Coupling}

\author{Christian W. Bauer}
\email{cwbauer@lbl.gov}
\affiliation{Physics Division, Lawrence Berkeley National Laboratory, Berkeley, CA 94720, USA}

\author{Dorota M. Grabowska}
\email{dorota.grabowska@cern.ch}
\affiliation{Theoretical Physics Department, CERN, 1211 Geneva 23, Switzerland}

\preprint{CERN-TH-2021-188}
\date{\today}

\begin{abstract}
We derive a representation for a lattice U(1) gauge theory with exponential convergence in the number of states used to represent each lattice site that is applicable at all values of the coupling.
At large coupling, this representation is equivalent to the Kogut-Susskind electric representation, which is known to provide a good description in this region. 
At small coupling, our approach adjusts the maximum magnetic field that is represented in the digitization as in this regime the low-lying eigenstates become strongly peaked around zero magnetic field. 
Additionally, we choose a representation of the electric component of the Hamiltonian that gives minimal violation of the canonical commutation relation when acting upon low-lying eigenstates, motivated by the Nyquist–Shannon sampling theorem.
For (2+1) dimensions with 4 lattice sites the expectation value of the plaquette operator can be calculated with only 7 states per lattice site with per-mille level accuracy for all values of the coupling constant. 
\end{abstract}

\maketitle
The Standard Model of Particle Physics, encapsulating the vast majority of our understanding of the fundamental nature of our Universe, is at its core a gauge theory. 
Much of the richness of its phenomenology can be traced back to the complicated interplay of its various gauged interactions. 
While massive theoretical and algorithmic developments in classical computing have allowed us to probe many of these aspects, there remain a plethora of open questions that do not seem amenable to these methods. 
With a fundamentally different computational strategy, quantum computers hold the promise to simulate the dynamics of quantum field theories from first principles, allowing access to ab-initio predictions of observables that are inaccessible using existing techniques on classical computers. 
In order to harness the full potential of quantum computers, an efficient implementation of the Hamiltonian of gauge theories on quantum processors is a mandatory first step. 
This is no simple task due to the redundancies inherent to any gauge theory, as well as the finite number of degrees of freedom inherent to any simulation.
For a review of various approaches, both analog and digital, see~\cite{Wiese:2013uua, Zohar:2015hwa, Dalmonte:2016alw, Aidelsburger:2021mia, Banuls:2019bmf, Zohar:2021nyc, Klco:2021lap}.
 
A consequence of gauge invariance is that the number of physical states, \textit{i.e.} those that obey Gauss' Law, is exponentially smaller than the number of states in the full Hilbert space. 
While techniques for enforcing gauge invariance have been developed that do not restrict to physical states (see for example \cite{Halimeh:2019svu, Lamm:2020jwv, Tran:2020azk, PhysRevLett.107.275301, PhysRevLett.109.125302,  Banerjee:2012pg, Kasper:2020owz, Hauke2013, PhysRevX.3.041018,  Kuhn2014, Stannigel:2013zka}), limited quantum resources make it preferable to define the Hamiltonian purely in terms of physical states. 
Additionally, in order to be implemented onto a digital quantum computer,  the physical Hilbert space must be finite. 
This requires not only choosing a truncation and digitization scheme, but also determining a finite-dimensional representation of the various operator components of the Hamiltonian. 
These schemes and representation must be chosen such that the discrete Hamiltonian reproduces the physics of the continuum theory with a sufficiently high fidelity and with quantifiable errors. 
There has been much effort in developing various methods and formulations, including magnetic or dual basis representations \cite{Kaplan:2018vnj,Unmuth-Yockey:2018xak, Haase:2020kaj, Bender:2020ztu}, prepotentials with a  basis of loop, string and hadron excitations \cite{Anishetty:2009nh,  Raychowdhury:2018osk, Raychowdhury:2019iki, Anishetty:2009ai, Manu2010, Manu2011, Raychowdhury:2013rwa},  discrete subgroups and  group space decimation \cite{Zohar2015, Zohar2017, Alexandru:2019nsa, Ji2020}, mesh digitization \cite{Hackett2019}, light-front formulations \cite{kreshchuk2020quantum, Kreshchuk2021} and orbifold lattice methods \cite{Buser:2020cvn, Kaplan:2002wv}. 
For work on experimental realizations, see~\cite{Muschik:2016tws, Kokail:2018eiw, Schweizer:2019lwx, mil2020scalable,  Yang:2020yer, Riechert:2021ink}.

For this work, we focus on Abelian lattice theories, particularly \uone lattice gauge theories . One well-known implementation of such a theory is the Kogut-Susskind (KS) Hamiltonian \cite{PhysRevD.11.395, RevModPhys.51.659, Robson:1980nt, LIGTERINK2000983c,  PhysRevD.31.2020}. 
In this formulation, the Hamiltonian is defined in terms of integer-valued electric fields, plus plaquettes that act as lowering operators, due to the canonical commutation relations. As this formulation is naturally written in the electric basis, it is easy to truncate the theory by truncating the electric field values. While the KS formulation is a general formalism for an untruncated theory in the electric-field basis, in this letter, we generally refer to the truncated version as the KS representation. 
This representation gives a highly efficient and accurate description in the limit of strong coupling, as in this regime the electric fluctuations are small and so the eigenstates of the electric Hamiltonian are close to the eigenstates of the full Hamiltonian.

On the other hand, the electric basis does not provide for an efficient representation of a \uone gauge theory when approaching the continuum limit. %
For gauge theories in two or less spatial dimensions, or asymptotically-free theories in three spatial dimensions, the continuum limit corresponds to the weak-coupling limit. 
As the continuum is approached and the coupling constant decreases, the widths of the eigenstate wave functions in the electric basis increase, such that a truncation at a fixed value of the electric field becomes inadequate. 
On the other hand, the support of the wave function in the magnetic basis decreases as the coupling decreases, indicating that these gauge theories are more efficiently represented in the magnetic basis.

In this work we derive a new digitized representation of a \uone gauge theory that that is efficient, regardless of the strength of the gauge coupling. 
The implementation of this representation proceeds in two steps. The first is determining the optimal digitization and truncation of the magnetic field values, as we always work in the magnetic basis. 
The second is determining the representation of the electric Hamiltonian in the magnetic basis. 
The main focus of this work is motivating a simple and analytic expression for the magnetic field digitization, as well as a  choice for the electric Hamiltonian that allows for a maximally faithful representation of the lowest-lying states.

This representation has several properties that are highly advantageous for quantum simulation. %
Primarily, this representation is extremely resource-efficient and is able to reproduce the lowest-lying eigenstates of the system with exponential precision. 
This efficiency is two-fold. The first is that the required number of states per lattice site is quite small for the degree of precision achieved. 
The second is that due to analytic expressions for optimal truncation values, the need for a computationally and time-intensive optimization routine is eliminated. 
While we leave a detailed study for future work, we believe that the representation presented here can be implemented onto qubits with minimal modification. 
These properties indicate that this representation is well-tailored for working near the continuum limit while utilizing quantum hardware. 
Additionally, the representation works regardless of the strength of the coupling; in fact, at large coupling it is related to the well-known KS formulation via a simple Fourier transform. 

Magnetic-basis formulations have previously been considered in, for example~\cite{Kaplan:2018vnj,Haase:2020kaj}, with \cite{Haase:2020kaj} focusing on creating a resource-efficient representation at weak coupling. 
After presenting the derivation of our representation, we will provide a brief discussion comparing to this work. 
More details can be found in the Supplemental Material.

The pure gauge part of a \uone gauge theory is given by the Hamiltonian
\begin{align}
    H = \frac{1}{2} \int \df^d x \left[\vec E(x)^2 +  B(x)^2 \right]
    \,.
\end{align}
where, for simplicity, we will work in (2+1) dimensions, and only comment about (3+1) dimensions in the end. Here the electric and magnetic field are related to the vector potential by
\begin{align}
    \vec E(x) = \frac{\df \vec A(x)}{\df t} \, , \qquad 
    B(x) = \vec\nabla \times \vec A(x)
    \,,
\end{align}
and we work in the $A_0(x) = 0$ gauge. Note that the curl of a vector field is a 2-form, which in (2+1) dimensions is dual to a scalar. 

Gauge invariance implies Gauss' law
\begin{align}
    \left[ \vec\nabla \cdot \vec E - \rho \right] \ket{\Psi} = 0
    \,,
\end{align}
giving a constraint on physical states $\ket{\Psi}$.
This constraint can be solved by writing~\cite{Drell:1978hr, Bender:2020ztu}
\begin{align}
    \vec E = \vec E^L + \vec E^T   
    \,,
\end{align}
where the longitudinal and transverse fields can be written as
\begin{align}
    \vec\nabla \cdot \vec E^L = \rho\,, \qquad \vec E^T = \vec \nabla \times R
    \,.
\end{align}
Here $\rho$ denotes the charge density, and $R$ is again a two-form.
Thus, in the absence of electric charges, the Hamiltonian can be written in terms of the two scalar quantities, $R$ and $B$, which satisfy the commutation relations
\begin{align}
\label{eq:commutation}
\left[ B(x), R(y) \right] = i \delta(x - y)
\,.
\end{align}
The field $R$ (which was called $L$ in the original work~\cite{Drell:1978hr}) was coined a ``rotor'' field in~\cite{Haase:2020kaj}. 

This Hamiltonian can be put directly on the lattice via two different formulations. 
These two formulations, corresponding to non-compact and compact \uone gauge groups, have the same continuum limit, but noticeably different behavior at finite lattice spacing,
Using dimensionless variables and rescaling $A \to A / g$, $E \to g E$, the Hamiltonian for a non-compact U(1) group is given
\begin{align}
    H^{\rm NC} & = \frac{1}{2a} \sum_p \left[ g^2 (\vec\nabla \times R_p)^2 + \frac{1}{g^2} B_p^2 \right]
    \nonumber\\
    & \equiv H_E + H_B^{\rm NC}
    \,.
\end{align}
where $a$ is the latice spacing. Here $g$ denotes the bare lattice coupling, whose relation to the physical observables in the continuum limit depends on parameters of the lattice. The sum is over plaquettes (with the subscript $p$ on the operators acting as an index) and  $\vec\nabla \times R_p$ is the lattice curl, defined in~\cite{Drell:1978hr}. 
Alternatively, the gauge field can be compactified onto a circle, leading to the compact version of the \uone gauge theory with $H^{\rm C} = H_E + H_B^{\rm C}$. 
This changes the magnetic Hamiltonian, whose compact form is given by
\begin{align}
    H_B^{\rm C} & = \frac{1}{a} \sum_p \frac{1}{g^2} (1 - \cos B_p)
    \nonumber\\
    & = \frac{1}{2a} \sum_p \frac{1}{g^2} (2 - P_p - P^\dagger_p)
    \,,
\end{align}
where the plaquette operator $P_p$ is defined as
\begin{align}
P_p = e^{i B_p}
\,.
\label{eq:MagOpeDef}
\end{align}
Note that the more familiar but analogous form of the plaquette operator is the product over the gauge links that are required when constructing gauge invariant interactions of gauge fields and fermions.
The commutation relations on the lattice are now given by
\begin{align}
\label{eq:commLatt}
    [B_{p}, R_{p'}] = i \delta_{p, p'}
    \,.
\end{align}
This compact representation is the KS Hamiltonian. 
In the following, we will usually work with the compact version of the Hamiltonian, but our final results will be applicable to the non-compact Hamiltonian as well. 

Setting the convention to denote operators by upper case letters and their eigenvalues by the corresponding lower case ones, notice that the compact nature of the magnetic field immediately leads to an integer spectrum in the rotor fields
\begin{align}
    R_p \ket{r_p} = r_p \ket{r_p}\,, \qquad r_p \in \mathbb{Z}
    \,.
\end{align}
In fact, from the commutation relation \eqref{eq:commLatt} one can easily verify that the plaquette operators acts as a lowering operator
\begin{align}
\label{eq:Rladder}
P_p \ket{r_p} = \ket{r_p-1}
\,.
\end{align}
Thus, the KS Hamiltonian is naturally represented in the electric (rotor) basis. 
One can switch to the magnetic basis through a Fourier transform
\begin{align}
\ket{b_p} = \sum_{r=-\infty}^\infty e^{i b_p r_p} \ket{r_p}
\, ,
\end{align}
which immediately demonstrates the compact nature of the magnetic states $\ket{b_p + 2\pi} = \ket{b_p}$.

In order to represent this field theory on digital (quantum) devices, the magnetic field, which is currently continuous, needs to be digitized. 
A standard way is to represent the magnetic field through $2L+1$ discrete equidistant points symmetric between $-\pi$ and $\pi$, which turns the continuous U(1) gauge group into a discrete ${\rm Z}(2L+1)$ gauge group, and introduces a spacing in the magnetic field given by $\delta b = 2 \pi / (2L+1)$. 
Note that by choosing an odd number of points one can cover $b = 0$ while keeping the representation symmetric. 
This digitization in turn introduces a maximum value $r_{\max}$ in the electric basis, such that $-L < r < L$. 

This representation is best suited to represent the theory at large values of the coupling, where the system is dominated by the electric Hamiltonian and eigenstates have most of their support at low values of $r$. However, it is very inefficient at small coupling.
This can be understood from the fact that at small coupling, the magnetic Hamiltonian dominates and the lowest-lying eigenstates therefore have support only in a narrow region around $b = 0$.
To accurately represent the sharply peaked wavefunctions requires a small value of $\delta b$, which in turn necessities large $L$, making the representation very costly.

A more efficient formulation can be found by working directly in the magnetic basis and digitizing the magnetic field values directly. 
This requires two key points to be addressed. The first is choosing a value for the maximal $b_p$ included in the digitization. As the magnetic Hamiltonian is diagonal in the magnetic basis, there is a natural `best choice' for its representation once the maximum $b_p$ is fixed. The second is determining the  representation of the electric Hamiltonian, which is no longer diagonal.

For the non-compact theory, the width of the wave function in the electric and magnetic basis ($w_R$ and $w_B$, respectively) scale with the coupling constant as
\begin{align}
    w_R \sim 1 / g\,, \qquad w_B \sim g
    \,.
\end{align}
For the compact theory, this approximation still holds for small coupling while at large coupling, the wavefunctions in the magnetic basis have support for the full space, $b_p \in [-\pi, \pi]$. 
An efficient digitization needs to sample values of $b$ and $r$ in the region of their support.
For small values of $g$ this requires the sampling of the magnetic field in the region
\begin{align}
\label{eq:BMax}
|b| < b_{\rm max} \sim g
\,,
\end{align}
while in the compact theory at large values of $g$ the sampling needs to be over the full range $|b| < \pi$.
In other words, \eqref{eq:BMax} \emph{defines} the region of support for the magnetic field, and one then uses $2\ell+1$ discrete values within this range to digitize the field value. 
Thus, we sample the magnetic field at each plaquette at the values
\begin{align}
    b_p^{(k)} = -b_{\max} + k \, \delta b\,, \qquad \delta b = \frac{b_{\max}}{\ell}
    \,,
\end{align}
with $0 \leq k \leq 2\ell$. 
This implies that the digitized conjugate rotor field satisfies
\begin{align}
    r_p^{(k)} = -r_{\rm max} + \left( k+\frac{1}{2} \right) \, \delta r 
    \,,
\end{align}
with with $0 \leq k \leq 2\ell$ and
\begin{align}
    \delta r = \frac{2\pi}{\delta b (2\ell+1)}\,, \qquad r_{\rm max} = \frac{\pi}{\delta b} 
    \,.
\end{align}
Note that for general $b_{\rm max}$ the spacing of the rotor fields $\delta r$ is no longer equal to 1, unlike in the compact undigitized lattice theory. This deviation away from $\delta r = 1$ is key to having an efficient and accurate representation.
As we will discuss below, in the limit $\ell \to \infty$ we recover the usual relation $\delta b = 2 \pi / (2\ell+1)$ such that $\delta r = 1$. 

Before addressing how to find the optimal value for $b_{\rm max}$, it is first necessary to discuss the optimal representation of the electric Hamiltonian in the magnetic basis.
Multiple representations of the electric Hamiltonian in the magnetic basis are possible, and all of them should belong to the same universality class. 
However, these representations differ in their resource requirements and also the degree of error they accumulate at finite lattice coupling. 
One choice is to use a finite difference representation of the relation $R = - i \partial_{b}$.
Another representation was provided in \cite{Haase:2020kaj}, which we review in the Supplemental Material. 

However, for our representation, we choose to follow the example of \cite{Klco:2018zqz} and represent the electric Hamiltonian by its exact eigenvalues and use a Fourier transform to put it into the magnetic basis. 
Recall that for conjugate variables, the eigenstates of the rotor operator are related by Fourier transform to the eigenstates of the magnetic field operator
\begin{align}
\label{eq:FT}
 \ket{r_p^{(k)}} &= \frac{1}{\sqrt{2\ell+1}} \sum_{k'=0}^{2\ell} e^{i \frac{2 \pi}{2\ell+1} (k-\ell) (k'-\ell-\frac12)} \ket{b_p^{(k')}} \nonumber \\
 &\equiv \sum_{k' = 0}^{2\ell} \left(\text{FT}\right)_{k k'}\ket{b_p^{(k')}}
 \,.
\end{align}
One can therefore represent each rotor via
\begin{align}
\label{eq:DigRotor}
\bra{b_p^{(k)}} R_p\ket{b_{p'}^{(k')}} = \sum_{n = 0}^{2\ell}  r_p^{(n)}\left(\text{FT}\right)^{-1}_{kn} \left(\text{FT}\right)_{nk'}\delta_{pp'}
\,.
\end{align}
The rest of this letter will be dedicated to demonstrating that, with an appropriate choice of $b_{\rm max}$, this representation is an extremely efficient and accurate choice for all values of the gauge coupling. 
Note that the magnetic field operator in the magnetic basis is simply
\begin{align}
\label{eq:DigMag}
\bra{b_p^{(k)}} B_p\ket{b_{p'}^{(k')}} = b_p^{(k)}\delta_{kk'}\delta_{pp'}
\,.
\end{align}

Having chosen the representation of the operators, the problem at hand is now how to choose the optimal value of $b^{(p)}_{\rm max}$ for a given value of $\ell$ such that the digitized theory is as close as possible to the continuum theory. 
It was shown in~\cite{Macridin:2018gdw,Macridin:2018oli} that the condition that the canonical commutation relation be minimally violated gives the optimal digitization for the quantum harmonic oscillator, with corrections suppressed exponentially in the dimension of the Hilbert space. We apply a very similar condition to find the optimal value of $b_{\rm max}$ for both the non-compact and compact theory.\footnote{One might worry that it is not appropriate to apply this criteria to the compact theory. However, at small coupling the compact and non-compact theory are quite similar. At large coupling, $b_{\rm max}$ is constrained by the maximal range of the magnetic field space itself and there is no need for truncation criteria.
In the Supplemental material we show that this condition is quite useful for a weakly-coupled compact version of the harmonic oscillator.} We choose this criteria since the canonical commutation relations are what defines the undigitized theory. Ensuring that they are minimally violated is crucial for producing a faithful representation and replicating the theory's fundamental features.
We define the `canonical commutator expectation value' as
\begin{align}
    \left\langle C^{(\ell)}_p \right\rangle[b_{\rm max}] \equiv 1 + i \bra{\Omega^{(\ell)}} \left[ B_p, R_p \right] \ket{\Omega^{(\ell)}}
    \,,
\end{align}
where the ground state and the spectrum of the operators also depend on the value of $b_{\rm max}$, $g$ and $\ell$. We now define $b_{\rm max}$ as the value that minimized $\langle C^{(\ell)}_p \rangle[b_{\rm max}]$,
\begin{align}
\label{eq:argmin}
 b^{(p)}_{\rm max}(g,\ell) = {\rm argmin}\left[ \langle C^{(\ell)}_p \rangle[b_{\rm max}]\right]
    \,,
\end{align}
and we indicated the dependence of $b_{\rm max}$ on the values of $g$ and $\ell$.
Note that in general the condition in~\eqref{eq:argmin} gives a different value of $b_{\rm max}$ for each plaquette. 
However these differences are quite small and so final analytic expression  utilizing uses a universal $b_{\rm max}$.

For a quantum harmonic oscillator (QHO), the condition that the canonical commutation relation be minimally violated has been previously used to derive analytically the optimal value of $b_{\rm max}(\ell) = 2 \ell \sqrt{\pi / (4\ell+2)}$ \cite{Macridin:2018gdw,Macridin:2018oli}. 
This analysis can be repeated for the non-compact \uone theory in the following way. 
First, notice that the non-compact \uone Hamiltonian is a three-dimensional QHO which can be reduced to three one-dimensional QHOs by neglecting terms that couple different lattice sites together. 
With this as-of-yet unjustified simplification, and taking into account relative factors of two and $g$, the optimal value of $b_{\rm max}$ is found to be
\begin{align}
\label{eq:bmaxanalyticalNC}
b^{\rm NC}_{\rm max}(g,\ell) = g \ell \sqrt{\frac{\sqrt{8}\pi}{2\ell+1}}
\,.
\end{align}
This expression is modified for the compact theory, due to the finite maximum range of the magnetic field. In this case, $b_{\rm max}$ is given by
\begin{align}
\label{eq:bmaxanalyticalC}
b^{\rm C}_{\rm max}(g,\ell) = \text{min}\left[b^{\rm NC}_{\rm max},\frac{2\pi \ell}{2\ell+1}\right]
\,,
\end{align}
where the second expression is the $b_{\rm max}$ value in the KS formulation. At sufficiently large values of $\ell$, $b_\text{max}^{C}$ will always be equal to $2\pi \ell/(2\ell+1)$ and therefore $\delta r = 1$.

With this final step, we now have a fully defined representation of a $2+1$ dimensional \uone lattice gauge theory that is valid at all values of the coupling. 
Before presenting several numerical tests of the proposal, let us quickly present the proposal in a unified, step-by-step manner.
\begin{enumerate}
\item Determine the desired (odd) number of states, $n$, per lattice site, the value of the bare gauge coupling and whether to use the compact or non-compact theory. Recall $\ell$ is related to $n$ via $n = 2\ell+1$.
\item Use \eqref{eq:bmaxanalyticalC} or \eqref{eq:bmaxanalyticalNC} to determine $b_{\rm max}$, depending on whether the desired gauge groups is compact or non-compact.
\item With $b_{\rm max}$ and $\ell$ in hand, the total Hamiltonian can be written down, using the operator definitions given in \eqref{eq:DigRotor} and \eqref{eq:DigMag}.
\end{enumerate}
We emphasis again that this procedure works well at any value of the gauge coupling and also for both the non-compact and the compact formulations. 

We conclude this letter by presenting two numerical tests of this formulation. 
In particular, we focus on the smallest possible system in $2+1$ dimensions, namely four lattice sites and periodic identification of the boundaries. Imposing Gauss's law by hand and constraining to the trivial topological sector, the degrees of freedom are three rotors and three plaquettes. This system was previously derived and studied in \cite{Haase:2020kaj} for the compact gauge group. 

The Hilbert space of this system is spanned by three magnetic fields, which we choose to denote as
\begin{align}
\ket{b^{(\bm k)}} &\equiv \underset{p}\otimes\ket{b_p^{(k_p)}}. \nonumber \\
&=\ket{b_1^{(k_1)}\,b_2^{(k_2)}\,b_3^{(k_3)}}
\,,
\end{align}
where $\bm k$ is the vector of state labels for the magnetic operators. The magnetic Hamiltonian for the compact theory is diagonal with
\begin{align}
\label{eq:HBFinal}
 & \bra{b^{(\bm{k})}} H^\text{NC}_B \ket{b^{(\bm{k'})}} 
    \\
    & \qquad = \frac{1}{a} \frac{1}{g^2} \left(4 - \sum_{p=1}^3\cos b_p^{(k_p)} - \cos \left[\sum_{p=1}^3 b_p^{(k_p)}\right]\right)\delta_{\bm k \bm k'}\nonumber
    \,,
\end{align}
while for the non-compact theory one replaces each $\cos(b)$ by $1 - b^2 / 2$.
The matrix elements of the electric Hamiltonian are given by
\begin{align}
\label{eq:HEFinal}
    & \bra{b^{(\bm{k})}} H_E \ket{b^{(\bm{k'})}} =- \frac{2g^2}{a} \sum_{n_i=0}^{2\ell} \left(\text{\textbf{FT}}\right)^{-1}_{\bm{k}\bm{n}}\left(\text{\textbf{FT}}\right)_{\bm{n}\bm{k}'}
    \nonumber\\
    & \qquad\quad \times\left(
    r_1^{(n_1)}\left(r_2^{(n_2)}+r_3^{(n_3)}\right)-\sum_{p=1}^3 \left(r_p^{(n_p)}\right)^2 \right)
    \,,
\end{align}
where we have used the notation $\left(\text{\textbf{FT}}\right)_{\bm{k}\bm{k}'}= \prod_i \left(\text{FT}\right)_{k_ik_i'}$. A diagram of this system is shown in Fig.~\ref{fig:RotorPic}
\begin{figure}
  \centering
  \includegraphics[width=0.45\textwidth]{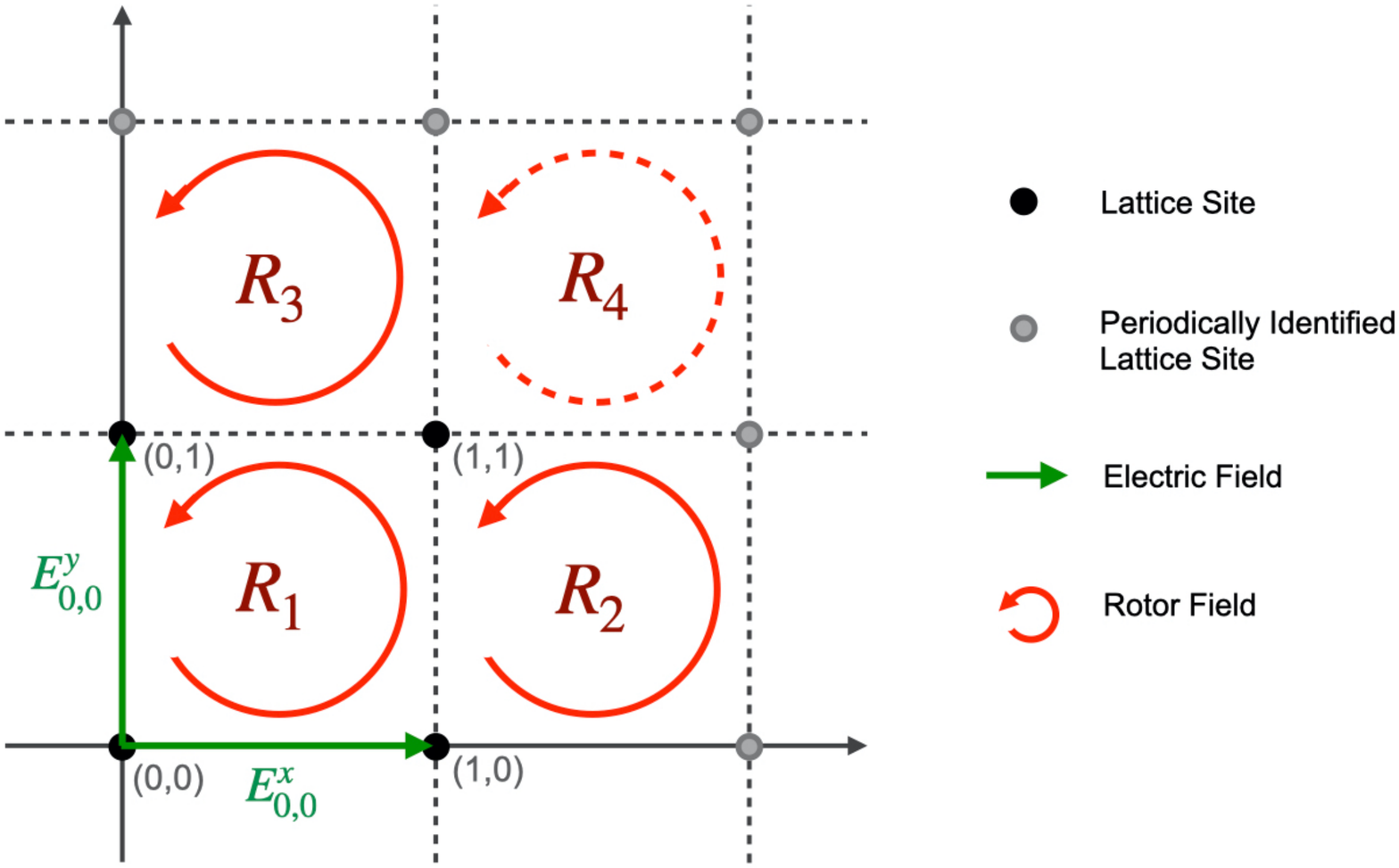}
  \caption{The smallest non-trivial system in two spatial dimensions, which four lattice sites and  boundaries that are periodically identified. The theory can be formulated either in terms of electric fields and plaquettes or in terms of rotors and plaquettes. One plaquette can always be removed.
  } 
\label{fig:RotorPic} 
\end{figure}

\begin{figure}
  \centering
  \includegraphics[width=0.48\textwidth]{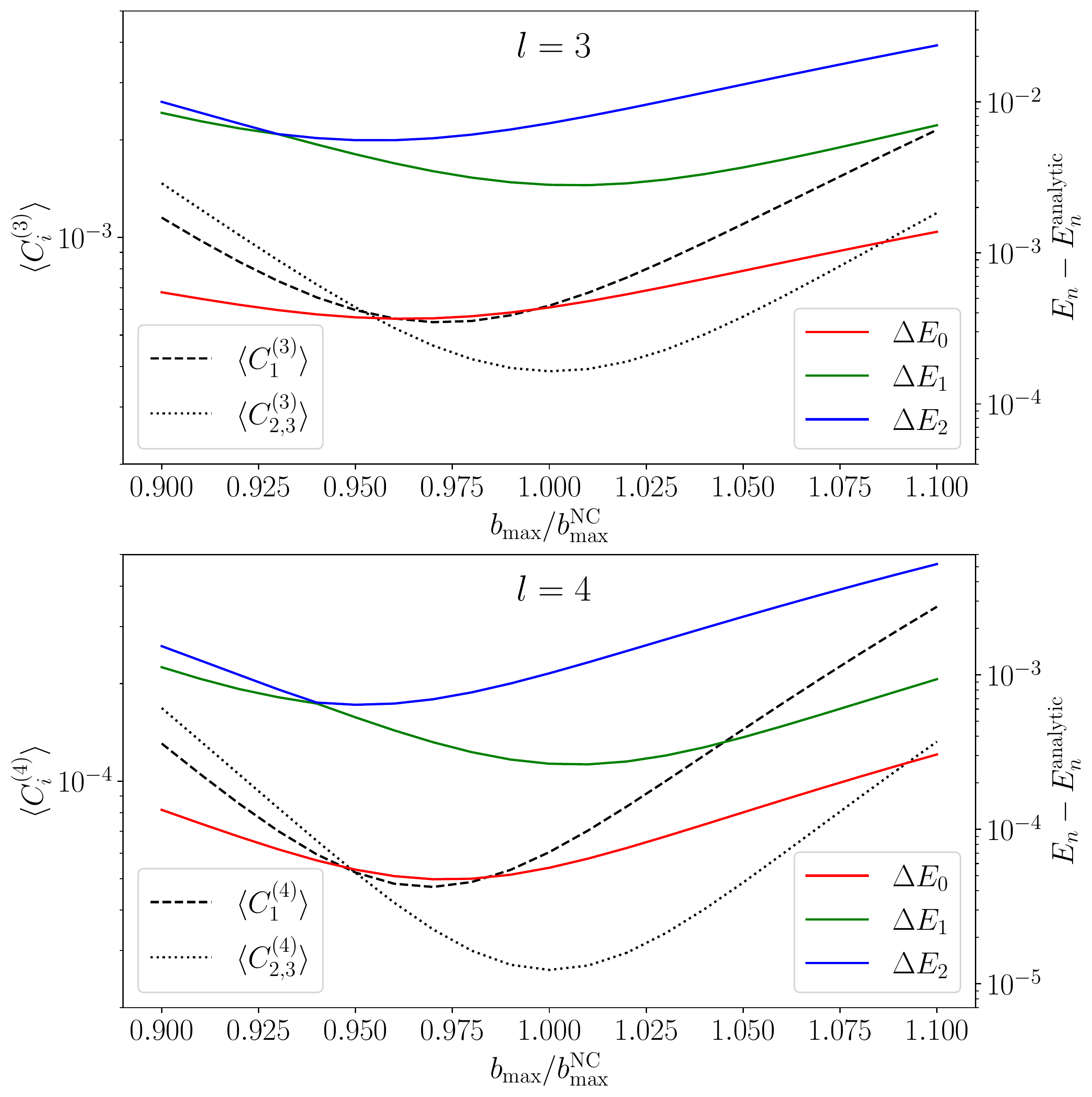}
  \caption{Dependence of the non-compact theory on the value of $b_{\rm max}$ for two values of $\ell$. The dashed and dotted lines show the commutator expectation value for the first and second/third lattice plaquette as a function of $b_{\rm max} / b^{\rm NC}_{\rm max}$. In the solid lines we show the results for energy difference compared to the analytical result. All curves are minimized for values of $b_{\rm max} \approx b^{\rm NC}_{\rm max}$. As these plots are for the non-compact theory, they are independent of the value of $g$.
  } 
\label{fig:bOptCheck} 
\end{figure}

In Fig.~\ref{fig:bOptCheck} we show the dependence of the theory on the value of $b_{\rm max}$ for the non-compact theory. 
The values of the $\langle C_p \rangle[b_{\rm max}]$ are shown by the dashed and dotted lines. The solid lines show the difference of the energies of the first three eigenstates of the Hamiltonian when compared to the exact value in the continuum limit. 
One can see that all curves have a minimum at very similar locations, and that these minima are very close to the analytical value $b_{\rm max}^{\rm C}$ given in~\eqref{eq:bmaxanalyticalC}.
The method to solve the non-digitized theory is presented in the Supplemental Materials (Section~\ref{app:noncompactU1}).

\begin{figure}[h]
  \centering
  \includegraphics[width=0.48\textwidth]{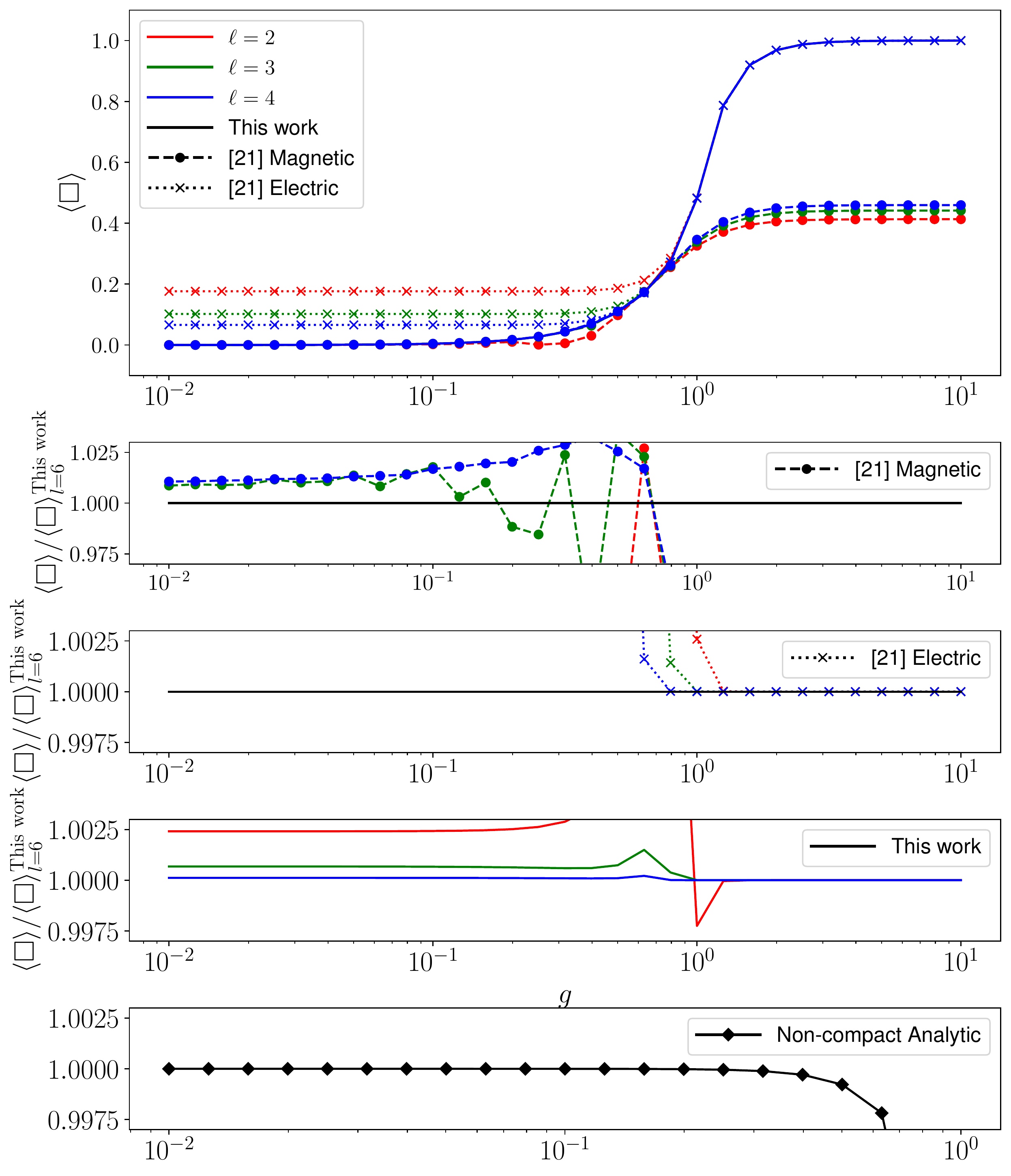}
  \caption{Expectation value of the plaquette operator for $\ell = 2$ (red), $\ell = 3$ (green), $\ell = 4$ (blue). The solid lines show the results of this work, while the dashed lines with circles and dotted lines with cross marks denote the results of~\cite{Haase:2020kaj} in the magnetic and electric basis, respectively. The ratios to the result of our work with $\ell = 6$ are shown below. In the bottom plot we also show the analytical solution of the non-compact theory, which should give the correct result at low values of $g$.}
\label{fig:ExpPlaquette} 
\end{figure}

As a second result, we present the expectation of the plaquette operator
\begin{align}
    \langle \Box \rangle = - \frac{g^2 a^2}{V} \bra{\Psi_0} H_B \ket{\Psi_0}
    \,,
\end{align}
where $\ket{\Psi_0}$ is the ground state of the theory and $V$ is the number of plaquettes in the system, 
which is four in this case.\footnote{Note that we eliminated one plaquette in the Hamiltonian as the effect of acting with this extra plaquette is the same as acting with the product of the Hermitian conjugate of the other three plaquettes.} 
This matrix element has been considered in the past~\cite{Paulson:2020zjd}, and allows for comparisons to~\cite{Haase:2020kaj}. 
The result is shown in Fig.~\ref{fig:ExpPlaquette}, where the solid lines correspond to the result of this work, while the dashed and dotted lines correspond to the results of~\cite{Haase:2020kaj} in the magnetic and electric basis, respectively. 
We can see that our results have very good convergence over the entire range of the coupling constant, and with $\ell = 3$ we have per-mille level accuracy for all values of $g$. 
As discussed before, the magnetic (electric) basis of~\cite{Haase:2020kaj} only works at small (large) couplings, and one can see that at small coupling the magnetic basis is only able to reach percent-level accuracy, even for larger $\ell$. 

To summarize, in this letter we presented a novel formulation of $(2+1)$-dimensional $U(1)$ lattice gauge theories. 
This formulation is able to reproduce the low-lying spectrum of the theory with per-mille or better accuracy while utilizing minimal resources. 
This formulation digitizes a Hamiltonian that only contains physical states (i.~e.~Gauss' law has been imposed \textit{a priori}), utilizing an analytic expression to estimate the optimal maximum field value based on the gauge coupling and the available resources for each lattice site.  
We believe that this procedure can be extended to larger systems in $(2+1)$-dimensions, though it becomes increasingly difficult to test this using classical resources, as the dimensionality of the Hilbert space scales with polynomial powers of $2\ell+1$. 
We leave the exploration of how well this formulation, particularly the analytic expression for $b_{\rm max}$, works in larger systems for future work involving quantum resources. 
Additionally, a similar procedure should be applicable to $(3+1)$ dimensions, though there arises an additional constraint that will complicate the procedure~\cite{Bender:2020ztu}. 
We close by noting that this work focused on a formulation of \uone gauge theories that was efficient in the overall dimension of the Hilbert space. 
An efficient implementation on digital quantum computers also requires a representation that can be implemented with efficient resources in quantum gates. 
While we believe our representation allows for such an efficient implementation, we leave a detailed study of the implementation in terms of quantum circuits in future work.

\begin{acknowledgments}
We thank Alessio Celi, Zohreh Davoudi, Hank Lamm, David B. Kaplan, Natalie Klco, Martin Savage, Jesse Stryker and Erez Zohar for helpful comments on the manuscript. CWB is supported by the U.S. Department of Energy (DOE), Office of Science under contract DE-AC02-05CH11231. In particular, support comes from Quantum Information Science Enabled Discovery (QuantISED) for High Energy Physics (KA2401032). DMG acknowledges the support of the CERN Quantum Technology Initiative. 
\end{acknowledgments}

\bibliographystyle{apsrev4-1}
\bibliography{myRefs}


\onecolumngrid
\clearpage
\newpage

\setcounter{secnumdepth}{\maxdimen}
\renewcommand{\theequation}{S\arabic{equation}}

\maketitle
\begin{center}
  \textbf{\large Supplemental Material to}\\
  \vspace{0.05in}
  \textit{ \large Efficient Representation for Simulating U(1) Gauge Theories on Digital Quantum Computers at All Values of the Coupling}\\ 
  \vspace{0.05in}
  Christian W. Bauer, Dorota M. Grabowska
\end{center}

\setcounter{equation}{0}
\setcounter{figure}{0}
\setcounter{table}{0}
\setcounter{section}{0}
\setcounter{footnote}{0}
\setcounter{page}{1}

\section{Review of the formulation of Haase et.~al.~\cite{Haase:2020kaj}}
In Ref.~\cite{Haase:2020kaj} an approach for an efficient representation at small coupling was advocated, which proceeded in two steps. 
First, the compact \uone theory was mapped onto a $Z(2L+1)$ theory, which digitizes the continuous $B$  field into $2L+1$ discrete values. 
This $Z(2L+1)$ theory has spacing 
\begin{align}
    \delta b(L) = \, \frac{2 \pi}{2L+1}
    \,,
\end{align}
while keeping $\delta r = 1$.
A second truncation was then performed that only kept the $(2 \ell +1)$ states with smallest absolute value of $b$. 
The truncation to $2\ell + 1$ states implies that all magnetic fields with value $b > b_{\rm max}$, with
\begin{align}
\label{eq:bmaxMusch}
    b_{\rm max}(\ell, L) = \ell \, \delta b = \frac{2 \pi \ell}{2L+1}
\end{align}
are neglected. 
The operator representation of the $Z(2L+1)$ is chosen such that the magnetic and electric representations are the Fourier transforms of  one another. Note that in~\cite{Haase:2020kaj} representations in both the electric and magnetic basis were given, and we review here only their magnetic basis. In the main paper, comparisons to both of their representations are presented.

While the representation of~\cite{Haase:2020kaj} reduces to the KS Hamiltonian for the choice $\ell = L$, as does the representation developed in this work for the choice $b_{\rm max}^{\rm NC}(\ell) = 2 \pi \ell / (2\ell+1)$, the two theories are fundamentally different for different values of $L$ or $b_{\rm max}$. The main difference lies in the representation of the electric Hamiltonian in the magnetic basis. 
In the work of~\cite{Haase:2020kaj} the truncated electric basis is obtained using the identity 
\begin{align}
    r &= \sum_{\nu=1}^{2L} f^s_\nu \sin \left( \frac{2 \pi \, \nu \, r}{2L+1} \right) 
     \nonumber\\
    r^2 &= \sum_{\nu=1}^{2L} f^c_\nu \cos \left( \frac{2 \pi \, \nu \, r}{2L+1} \right) + \frac{L(L+1)}{3}
    \,,
\end{align}
with
\begin{align}
    f^s_\nu &= \frac{(-1)^\nu+1}{2\pi} \left[ \psi_0\left(\frac{2L+1+\nu}{4L+2}\right) - \psi_0\left(\frac{\nu}{4L+2}\right)\right]
    \nonumber\\
    f^c_\nu &= \frac{(-1)^\nu}{4\pi^2} \left[ \psi_1\left(\frac{\nu}{4L+2}\right) - \psi_1\left(\frac{2L+1+\nu}{4L+2}\right)\right]
    \,,
\end{align}
which holds for integer values of $r$. 
The $\sin$ and $\cos$ functions can be related to exponentials using
\begin{align}
    \sin \left( \frac{2 \pi \, \nu \, r}{2L+1} \right) = \frac{e^{i \frac{2 \pi \, \nu \, r}{2L+1}} - e^{i \frac{2 \pi \, \nu \, r}{2L+1}}}{2i}
    \nonumber\\
    \cos \left( \frac{2 \pi \, \nu \, r}{2L+1} \right) = \frac{e^{i \frac{2 \pi \, \nu \, r}{2L+1}} + e^{i \frac{2 \pi \, \nu \, r}{2L+1}}}{2}
    \,,
\end{align}
Combining these relations with the operator relation (with $\ket{b^{(L+1)}} = \ket{b^{(-L)}}$)\footnote{Note that the integer values labeling the digitized magnetic field are now chosen to run over the values $-\ell < k < \ell$, rather than $0 < \ell < 2\ell$ as chosen in our approach.}
\begin{align}
    e^{i \frac{2 \pi \, \nu \, R_p}{2L+1}} \ket{b^{(k)}_p} = \ket{b^{(k+\nu)}_p}
    \,,
\end{align}
gives the expression for the rotor field operators in the magnetic basis
\begin{align}
    R_p &= \sum_{\nu=1}^{2L} f^s_\nu \frac{1}{2i} \sum_{k=-\ell}^{\ell} \left[ \ket{b_p^{(k)}} \bra{b_p^{(k+\nu)}} - \ket{b_p^{(k+\nu)}} \bra{b_p^{(k)}}\right]
    \nonumber\\
    R_p^2 &= \sum_{\nu=1}^{2L} f^s_\nu \frac{1}{2} \sum_{k=\ell}^{\ell} \left[ \ket{b_p^{(k)}} \bra{b_p^{(k+\nu)}} + \ket{b_p^{(k+\nu)}} \bra{b_p^{(k)}}\right] + \frac{L(L+1)}{3} \sum_{k=-\ell}^{\ell}\ket{b_p^{(k)}} \bra{b_p^{(k)}}
    \,.
\end{align}

As already mentioned, for $\ell = L$ this representation agrees with the one chosen in our work if the same value of $b_{\rm max} = 2 \pi \ell / (2\ell+1)$ is used. However, once truncated with $\ell < L$, the two representations no longer agree, even if one ensures that the $b_{\rm max}$ values agree. Note that as explained in~\cite{Haase:2020kaj}, the truncation is performed such that applying a lowering operator to the state $\ket{b^{(-\ell)}}$ simply annihilates the state, and does not give the state $\ket{b^{(\ell)}}$. 

Proceeding  with the construction of~\cite{Haase:2020kaj}, the discussion so far has determined the representation of the operators in the magnetic basis, given a value of $\ell$ chosen small enough that it can be simulated on a given choice of hardware. The only thing left to do is to choose a value $L$, which amounts to choosing a value of $\delta  b$ that allows for an efficient sampling of the states in the magnetic basis.  The work of~\cite{Haase:2020kaj} advocated to use a so-called sequence fidelity, defined as
\begin{align}
\label{eq:sequenceFid}
    F(l, L) = \sum_{\vec{k}=-\ell}^\ell \braket{\Psi_0(\ell,L)}{b^{(\vec{k})}}\braket{b^{(\vec{k})}}{\Psi_0(\ell+1,L)}
    \,.
\end{align}
For small values of the coupling constant, the sequence fidelity generally exhibits a local minimum when plotted as a function of $L$ (For $\ell = 2$ a local minimum is not present for all values of the coupling constant, but a kink in the dependence of the sequence fidelity when plotted as a function of $L$ is often still present.)
  \begin{figure}[!h]
    \centering
    \includegraphics[width=0.75\textwidth]{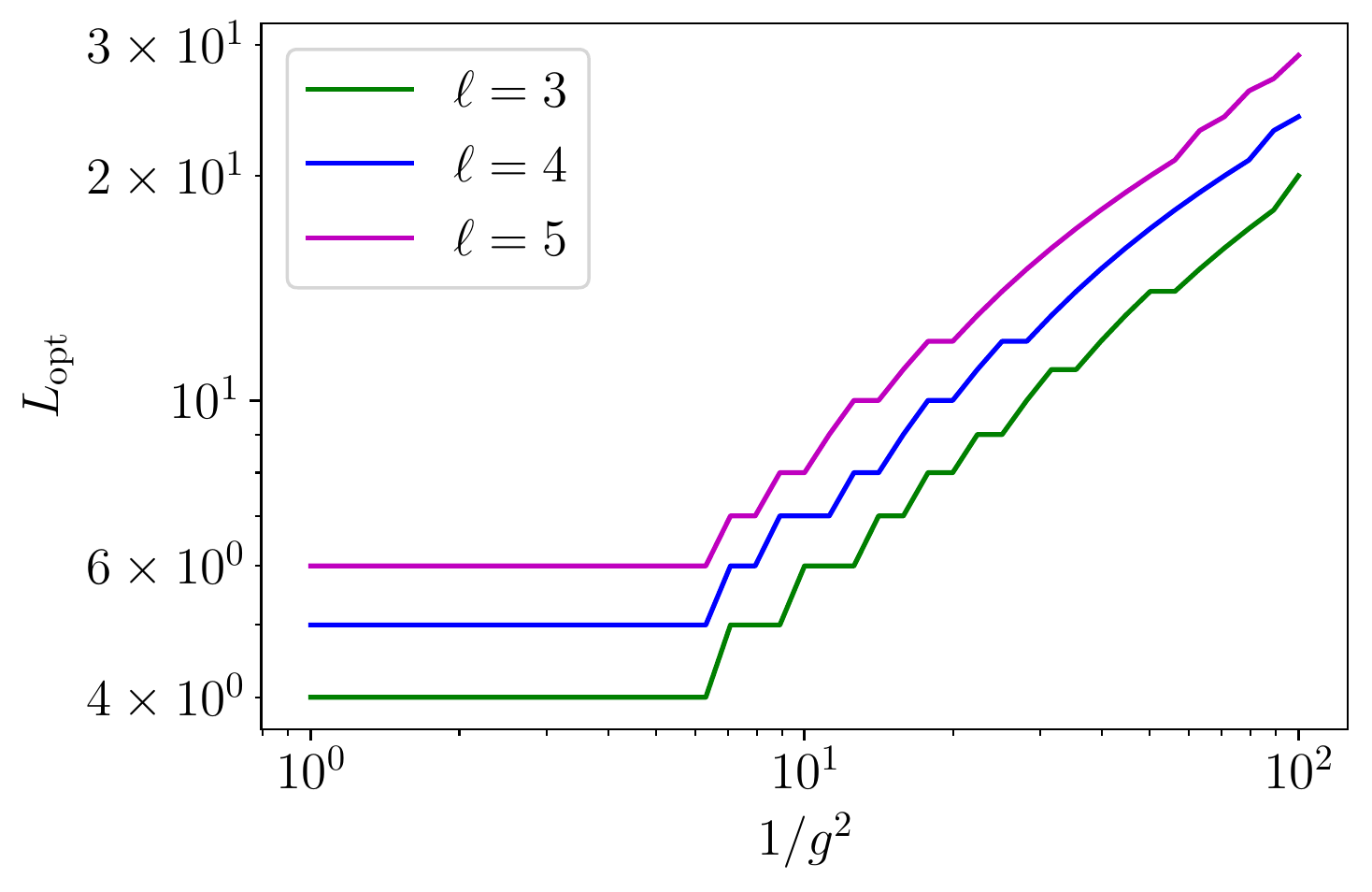}
    \caption{Dependence on the value of $L_{\rm opt}$ as defined via the sequence fidelity given in Eq.~\eqref{eq:sequenceFid} on the coupling constant $g$. We show the results for values $\ell  = 3, 4, 5$ in green, blue and magenta, respectively. For values above about $g \lesssim 0.37$ the value of $L_{\rm opt}$ starts to rise proportional to $g$. The plot looks similar to the Fig. 3(d) of~\cite{Haase:2020kaj}, but a detailed comparison is not possible due to the log-log nature of the plots.}
    \label{fig:Lopt}
  \end{figure}
The value of $L$ where this local minimum or kink occurs is the chosen value of $L$. For larger values of the coupling constant the local minimum disappears, and in this case~\cite{Haase:2020kaj} choose the value $L = \ell+1$. 
In Fig.~\ref{fig:Lopt} we show the results of the optimal value $L_{\rm opt}$ we obtain from this analysis. 
While the log-log plots make it difficult to compare our results exactly to the results in Fig.~3(d) of~\cite{Haase:2020kaj}, our results seem to be in general qualitative agreement. 
For a coupling of $g = 0.1$ the values of $L_{\rm opt}(\ell)$ we obtain are given in Table~\ref{tab:Lopt}. The values of $L_{\rm opt}$ we obtain for this value of $g$ are in agreement with the values that can be read off Fig.~9 of~\cite{Haase:2020kaj}. 
\begin{table}[h!]
    \centering
    \begin{tabular}{|c||c|c|c|c|c|c| c| c|}
        \hline
         $\ell$ & 2& 3 & 4 & 5 & 6& 7& 8 & 9\\\hline\hline
         $L_{\rm opt}$ & 14 & 20 & 24 & 29 & 32 & 36 & 39 & 42\\\hline
         $ b_{\rm max}(\ell, L_{\rm opt})$ & 0.43& 0.46 & 0.51 & .53 & 0.58 & 0.60 & 0.64 & 0.67 \\
         $ b_{\rm max}^{\rm C}(\ell)$ &  0.27 & 0.34 & 0.40 & 0.45 & 0.50 & 0.54 & 0.58 & 0.62\\\hline
    \end{tabular}
    \caption{The values of $L_{\rm opt}$ as determined from the sequence fidelity in Eq.~\eqref{eq:sequenceFid} for $g = 0.1$. The obtained values agree with the results that can be read off from Fig.~9 of ~\cite{Haase:2020kaj}. We also show the corresponding value of $b_{\rm max}(\ell, L_{\rm opt})$ that is obtained from Eq.~\eqref{eq:bmaxMusch} in comparison with the value obtained from the procedure described in this work}
    \label{tab:Lopt}
\end{table}

We finish this section with a few comments. First, as already mentioned, the magnetic representation of~\cite{Haase:2020kaj} chooses $L_{\rm opt}$ to be at least of size $\ell + 1$. This implies that it does not coincide for any value of $g$ or $\ell$ with the representation presented in this work. Furthermore, it also means that the magnetic representation is never equivalent to the electric basis, as it would be for the choice $L = \ell$. We believe that this is the main reason why the magnetic basis of~\cite{Haase:2020kaj} performs so poorly at large values of $g$. Second, choosing a value $L$ is equivalent to choosing a value of $\delta b(\ell, L_{\rm opt})$. As can be seen from Table~\ref{tab:Lopt} the resulting value of $\delta b$ differs from the analytical value used in our work, but the two values have the same overall trend, as expected. Finally, we reiterate that even choosing the exact value of $b_{\rm max}$ as was done in our work, the two representations differ. 

\section{A simple toy model: the harmonic oscillator}
\label{app:ToyModel}
In this Section, we present a toy model that can be solved analytically for both the compact and non-compact form. The main purpose of this toy model is to explore the error induced in the the compact formulation from using the analytic expression for $\bMax$. 

The toy model is a one-dimensional quantum harmonic oscillator (QHO) and we look at a non-compact as well as a compact formulation. 
The non-compact version is simply the standard QHO
\begin{align}
H &= \frac{g^2 \beta^2_p}{2}P^2 + \frac{\beta^2_x}{2g^2} X^2 \nonumber \\
&= H_P + H^{\rm NC}_X
\,,
\end{align}
where we have introduced a coupling $g$ to more closely resemble the full theory, as well as the parameters $\{\beta_x, \beta_p\}$ to allow for relative rescalings of the operators. The compact Hamiltonian is given by
\begin{align}
H &=  \frac{g^2 \beta^2_p}{2}P^2 + \frac{\beta^2_x}{2g^2}  \left(2 - 2 \cos X\right) \nonumber \\
&= H_P + H^{\rm C}_X
\,.
\end{align}
In both formulations, the operators $X$ and $P$ have the canonical commutation relation
\begin{align}
\left[X,P\right] = i
\,,
\end{align}
and we choose to work with the variables $\{X, P\}$ instead of $\{B, R\}$ to make the notation more closely resemble a harmonic oscillator.

\subsection{Analytical Solutions}
\label{app:AnalySol}
The analytical solution to the quantum harmonic oscillator is well known, and the eigenvalues and eigenstates are given by
\begin{align}
\label{eq:noncompactAnaltyicalToy}
    E_n &= \beta_x \beta_p\left(n + \frac{1}{2}\right)\, \nonumber \\
    \braket{x}{\phi_n} &=\sqrt{\frac{1}{2^n n!}}\left( \frac{1}{\pi g^2}\frac{\beta_x}{ \beta_p}\right)^{1/4} e^{-\frac{\beta_x}{\beta_p}\frac{x^2}{2g^2}} H_n\left[\sqrt{\frac{\beta_x}{\beta_p}}\frac{x}{g}\right]
    \,,
\end{align}
We have expressed the wave function in the basis spanned by eigenstates of the operator $X$
\begin{align}
    X\ket{x} = x \ket{x}\,, \qquad \braket{x}{x'} = \delta(x-x')
    \,.
\end{align}  
The compact version of the Harmonic oscillator is more complicated to solve analytically, but can be expressed in terms of Mathieu functions of the first kind. Mathieu functions are solutions to the characteristic second order differential equation
\begin{align}
\frac{d^2 y}{dz^2}+ \left(a - 2 q \cos 2 z\right)y = 0
\,.
\end{align}
There exist a specific class of these functions, generally called \textit{Matthieu Functions of the First Kind of Integral Order} that have either a period of $\pi$ or $2\pi$ and definite odd or even parity. 
These four functions only occur when the parameters $a$ and $q$ are suitably related. 
We follow the convention that the Matthieu functions that reduce to $\cos n z$ when $q =0$ are called ${\rm Ce}_n(z)$ and those that reduce to $\sin n z$ when $q =0$ are called ${\rm Se}_n(z)$. 
The corresponding values, which we call `characteristic numbers' of $a$ are denoted by $a_n$ for ${\rm Ce}_n(z)$ and $b_n$ for ${\rm Se}_n(z)$. 
The functions ${\rm Ce}_n(x, q)$ and ${\rm Se}_n(x, q)$ are also called cosine-elliptic and sine-elliptic functions. 
Summarizing, the periodicity, parity and characteristic numbers are shown in Table~\ref{tab:CharAB}. 
A property of these characteristic numbers is that for a given value of $n$ and scanning through different values of $q$, the different characteristic numbers cannot intersect, except for at $q =0$~\cite{arscott1964periodic}. 
This allows us to assign a state label to each energy level that is independent of the coupling constant. 
For negative $q$, which is the region of interest for this system, the ordering is as follows
\begin{align}
&a_0 < b_{2r} <a_{2r} \nonumber\\
&a_{2r-1}<b_{2r-1} \qquad r = 1, 2, 3, \dots
\end{align}
It is amusing to note that for positive $q$, $a_0 < b_n < a_n$ for any positive integer $n$.

\begin{table}[t]
    \centering
    \begin{tabular}{c|c|c|c|c}
    Function & n & Period & Parity &Characteristic Value\\ \hline \hline
$ce_n(z)$&\quad$ 0, 2, 4, \dots$\quad& $\pi$ & even & $a_n$
\\
$ce_n(z)$&\quad$1, 2, 3, \dots$\quad& $2\pi$ & even & $a_n$\\
$se_n(z)$&\quad$2, 4, 6, \dots$\quad& $\pi$ & odd & $b_n$  
\\
$se_n(z)$&\quad$1,2,3, \dots$\quad& $2\pi$ & odd & $b_n$
    \end{tabular}
    \caption{The period and parity of the four different Matthieu Functions of the First Kind of Integral Order. We also show the convention for denoting the corresponding value of $a$, for a given $q$, such that the functions have the desired periodicity.}
    \label{tab:CharAB}
\end{table}
For the toy model, all four types of Matthieu functions are allowable eigenstates. However, upon digitization of the theory, only certain eigenvalues appear, depending on the exact digitization of the eigenvalues of the $H_P$. 
If the eigenvalues of $H_P$ are symmetric around zero, only the states with period $\pi$ appear, while if the eigenvalues of $H_P$ are asymmetric around zero, only the states with period of $2\pi$ appear. 
These different choices in digitization are equivalent to setting different boundary conditions. 
Combining all of this, the solutions for the toy model can be classified by whether the eigenvalues of $H_P$ are symmetric or not. For the boundary condition that results in a symmetric digitization, the energies are given by
\begin{align}
E_n^{(S)} &=\frac{g^2 \beta_p^2}{8}\, f^{(S)}_{n}\left[q\right] + \frac{\beta_x^2}{g^2} \qquad \qquad q = -\frac{4}{g^4} \frac{\beta_x^2}{ \beta_p^2} \qquad \qquad n = 0, 1, 2, 3, \dots
\end{align}
where the functions $f_n^{(S)}$ are related the the Matthieu characteristic numbers via
\begin{align}
f^{(S)}_{n}[x] &= \left\{
\begin{array}{ll}
a_n[x] &\qquad n \, {\rm even}\\
b_{n+1}[x] & \qquad n \, {\rm odd}\\
\end{array}\right.
\,,
\end{align}
and the corresponding eigenfunctions are given by
\begin{align}
\braket{x}{\phi^{(S)}_{n}} &= \left\{
\begin{array}{ll}
N_n^{(S)}{\rm Ce}_{n}\left(\frac{x}{2}, q\right) &\qquad n \, {\rm even}\\
N_{n}^{(S)}{\rm Se}_{n+1}\left(\frac{x}{2}, q\right) &\qquad  n \, {\rm odd}\\
\end{array}\right.
\,.
\end{align}

For the boundary condition that gives rise to  asymmetric boundary conditions, the energies are given by
\begin{align}
E_n^{(A)} &=\frac{g^2 \beta_p^2}{8}\, f^{(A)}_{n}\left[q\right] + \frac{\beta_x^2}{g^2} \qquad \qquad q = -\frac{4}{g^4} \frac{\beta_x^2}{ \beta_p^2} \qquad \qquad n = 0, 1, 2, 3, \dots
\end{align}
where the functions $f_n^{(A)}$ are again related the the Matthieu characteristic numbers via
\begin{align}
f^{(A)}_{n}[x] &= \left\{
\begin{array}{ll}
a_{n+1}[x] & \qquad n \, {\rm even}\\
b_{n}[x] & \qquad  n \, {\rm odd}\\
\end{array}\right.
\,,
\end{align}
and the corresponding eigenfunctions are given by
\begin{align}
\braket{x}{\phi^{(A)}_{n}} &= \left\{
\begin{array}{ll}
N_{n}^{(A)}{\rm Ce}_{n+1}\left(\frac{x}{2}, q\right) & n \, {\rm even}\\
N_{n}^{(A)}{\rm Se}_{n}\left(\frac{x}{2}, q\right) & n \, {\rm odd}\\
\end{array}\right.
\,.
\end{align}
When comparing these analytical solution to the digitized theory, the periodic or antiperiodic bounndary conditions are imposed via the exact representation chosen for the $H_P$. We show the first few energy values and eigenstates in Table ~\ref{tab:EigMatthieu}.

\begin{table}[t]
\begin{subtable}{.5\linewidth}
\centering
\begin{tabular}{c|c|c|c|c}
 \quad$n$  \quad& \quad $f^{(S)}_n$  \quad & \quad $\phi^{(S)}_n$ \quad &  \quad $E^{(S)}_n( g = 0.5)$  \quad &  \quad $E^{(S)}_n( g = 2)$ \\ \hline \hline
0 &  $a_0$ & $\text{Ce}_0$& 0.492059& 0.23448\\
1 &  $b_2$ & $\text{Se}_2$& 1.45971& 2.2474\\
2 &  $a_2$ & $\text{Ce}_2$& 2.3936& 2.26291\\
3 &  $b_4$ & $\text{Se}_4$& 3.29152& 8.25104\\
4 &  $a_4$ & $\text{Ce}_4$& 4.15087& 8.25104
\end{tabular}
\captionsetup{width=.9\linewidth}
\caption{The first few eigenstates and their corresponding energies for the toy model, with boundary conditions that result in a symmetrically digitized $H_P$.}
\end{subtable}%
\begin{subtable}{.5\linewidth}
\centering
\begin{tabular}{c|c|c|c|c}
 \quad$n$  \quad& \quad $f^{(A)}_n$  \quad & \quad $\phi^{(A)}_n$ \quad &  \quad $E^{(A)}_n( g = 0.5)$  \quad &  \quad $E^{(A)}_n( g = 2)$ \\ \hline \hline
0 &  $a_1$ & $\text{Ce}_1$& 0.492059& 0.621214\\
1 &  $b_1$ & $\text{Se}_1$& 1.45974& 0.870971\\
2 &  $a_3$ & $\text{Ce}_3$& 2.3936& 4.75183\\
3 &  $b_3$ & $\text{Se}_3$& 3.29152& 4.75208\\
4 &  $a_5$ & $\text{Ce}_5$& 4.15087& 12.7507
\end{tabular}
\captionsetup{width=.9\linewidth}
\caption{The first few eigenstates and their corresponding energies for the toy model, with boundary conditions that result in an asymmetrically digitized $H_P$.}
\end{subtable} 
\caption{The first few eigenstates and their corresponding energies for the toy model. We provide the value of the energies for two different values of the coupling, $g = 0.5$ and $g = 2$, evaluated at $\beta_x = \beta_p = 1$. Notice that for small coupling, the energies are close the QHO with frequency equal to one.}
\label{tab:EigMatthieu}
\end{table}

The analytic solution for this class of toy models was also found in \cite{Bender:2020jgr} and we agree with their result if $\beta_x = 1$ and $\beta_p = 2$ and periodic boundary conditions are imposed.

\subsection{Digitization}
This toy model can also be solved by digitization. 
In this case, instead of having a continuous Hilbert space characterized by the basis states $\ket{x}$, one chooses a discrete set of states $\ket{x_i}$ with $\braket{x_k}{x_{k'}} = \delta_{kk'}$ and
\begin{align}
\label{eq:xTrunc}
   x^{(k)} = -x_{\max} + k \delta x\,, \qquad \delta x = \frac{x_{\max}}{\ell}
    \,.
\end{align}
In our convention, $k \in [0, 2\ell]$ and the Hilbert space therefore has dimension $2\ell +1$ and $\{x_\text{max}, \ell\}$ are parameters that can be freely chosen, though there is an optimal value of $x_\text{max}$ for each value of $\ell$.

The Hamiltonian $H_X$ is easily written in its digitized form
\begin{align}
    \bra{x_k} H_X^{\rm NC} \ket{x_{k'}} &= \frac{1}{2a} \frac{\beta_x^2}{g^2} \, x_k^2 \, \delta_{kk'} \nonumber\\
    \bra{x_k} H_X^{\rm C} \ket{x_{k'}} &= \frac{1}{2a} \frac{\beta_x^2}{g^2} \left(2-2\cos x_k \right) \delta_{kk'} \nonumber
    \,.
\end{align}
For the Hamiltonian $H_P$ there are several definitions possible. The first is to use the operator relation $P^2 = - d^2 / dx^2$ and to use a discretized second derivative to obtain
\begin{align}
    \bra{x_k} H_P^{(1)} \ket{x_{k'}} &= \frac{1}{2a} \frac{\beta_p^2g^2}{(\delta x)^2}\left[2\delta_{kk'} - \delta_{k,k'-1} - \delta_{k-1,k'} \right]
    \,,\nonumber\\
    & \equiv \frac{1}{2a} \frac{\beta_p^2g^2}{(\delta x)^2} \left[2 \mathbb{I}_{kk'} - T_{kk'} - T^\dagger_{kk'}  \right] 
    \,.
    \label{eq:HR1}
\end{align}
Here $\mathbb{I}_{kk'}$ and $T_{kk'}$ denote the identity and lowering operator, respectively
\begin{align}
    \mathbb{I} \ket{x_k} = \ket{x_k} \,, \qquad T \ket{x_k} = x_{k-1} 
    \,.
\end{align}
Note that the lowering operator can also be thought of as a translation operator in momentum space. In this representation, there is only one free parameter, the integer $L$. Since the digitized system is finite, boundary conditions must be imposed. For periodic boundary conditions and open boundary conditions, the lowering operator obeys, respectively,
\begin{align}
T^{\text{(PBC)}} \ket{x_0} = T_{2N} \qquad T^{\text{(OBC)}} \ket{x_0} = 0
\,.
\end{align}
However, as was shown in Ref.~\cite{Klco:2018zqz}, this definition of $H_P$ does an extremely poor job of reproduce the low-energy spectrum of the theory due to poor sampling of the region that has the largest support. 

An alternative representation is to use the `exact' eigenvalue method, advocated for in Ref.~\cite{Klco:2018zqz} and used in the main body of this manuscript. This method uses the fact that $H_P$ is diagonal in the conjugate space spanned by $\ket{p_k}$, where the two spaces are related by Fourier transform. The conjugate space satisfies
\begin{align}
p_k = -p_\text{max} + \left(k+\frac{1}{2}\right) \delta p 
\,,
\end{align}
where we have chosen a symmetric digitization of the momentum eigenvalues and
\begin{align}
\label{eq:pTrunc}
\delta p = \frac{2\pi}{\delta x (2\ell+1)} \qquad p_\text{max} = \frac{\pi}{\delta x}
\,.
\end{align}
The Hamiltonian is then given by
\begin{align}
 \bra{x_k} H_P^{(2)} \ket{x_{k'}} &= \frac{\beta_p^2 g2}{2a}\sum_{n} p^2_n \left(\text{FT}\right)^{-1}_{kn}\left(\text {FT}\right)_{nk'}
 \,.
\end{align}
In this representation, there are two free parameters, $x_\text{max}$ and $\ell$ and so we are free to set $x_\text{max}$ to an optimal value. 

\subsection{Choosing an optimal value of $\delta x$}
The optimal value of $\delta x$ is chosen such that the lowest-lying eigenvalues of the digitized theory are exponentially close to the eigenvalues of the continuum theory. For the non-compact theory, the optimal value can be motivated by looking at the eigenstates in the continuum theory. Note that any one-dimensional QHO can be rescaled such that
\begin{align}
H = \frac{1}{2}\tilde X^2 + \frac{1}{2}\tilde P^2
\,.
\end{align}
The eigenstates of this Hamiltonian are given by Eq.~\eqref{eq:noncompactAnaltyicalToy} with $g = \beta_x = \beta_y = 1$. 
For this choice of parameters the width of the wave function in position and momentum space are  the same, and one can therefore intuitively understand that the best digitization of this theory will have $\delta p = \delta x$. 
Combining this with \eqref{eq:xTrunc} and \eqref{eq:pTrunc} gives $\tilde{x}_\text{max}(\ell) = \ell \sqrt{2\pi/(2\ell+1)}$~\cite{Bauer:2021gup}.
Carefully undoing the re-scaling results in a expression for $x_{\text{max}}$ of
\begin{align}
\label{eq:MSAHCond}
\tilde{x}_\text{max}(\ell) = g \, \ell \, \sqrt{\frac{\beta_p}{\beta_x}} \sqrt{\frac{2\pi}{2 \ell +1}}
\,.
\end{align}

This intuitive argument was put on firm footing by Ref.~\cite{Macridin:2018gdw,Macridin:2018oli}, where it was shown that \eqref{eq:MSAHCond} can be derived by requiring that in the digitized theory the canonical commutation relation $[X, P] = i$ is minimally violated. With the choice of~\eqref{eq:MSAHCond} one can show that the commutation relations satisfy
\begin{align}
   \sum_{k'} \bra{x_k} [X, P] \ket{x_{k'}}\braket{x_{k'}}{\Psi_n} = i \braket{x_k}{\Psi_n} + {\cal O}(\epsilon_n)
   \,,
    \label{eq:commutatorerror}
\end{align}
where
\begin{align}
    \epsilon_n \sim \int_{x_{\rm max}(\ell)}^\infty \!\!\! \rm {\rm d} x \, \left| \braket{x}{\Psi_n} \right|^2
    \label{eq:epsdef}
    \,,
\end{align}
is of order the support of the wave function that lies beyond the maximal value $x_{\rm max}(\ell)$, which vanishes exponentially as $\ell$ is increased. 
Note, however, that $\epsilon_n$ grows with $n$, which implies that the correct commutation relations are only observed for the lowest lying states for which \eqref{eq:epsdef} is sufficiently small. 
Since the eigenvalues of the Hamiltonian are derived from the commutation relations, one immediately finds that the digitized model gives energy eigenvalues $E_n^{(\ell)}$ which approach the correct energy eigenvalues exponentially as $\ell$ is increased
\begin{align}
    \epsilon_n(L) = |E_n - E^{(\ell)}_n| \sim a_n e^{-\left(2\ell+1\right)}
    \,.
\end{align}

For the compact version of the QHO the formal proofs do not apply directly. 
However, given that the value given in~\eqref{eq:MSAHCond} gives a very good approximation to the commutation relation for all values of $n$ for which the support of the wave function above $x_{\rm max}$ is small, it gives a good approximation to any linear combination of these wave functions. 
One therefore should expect this relation to also work on the eigenstates of a different theory, as long as the support of the lowest lying wave functions has a very similar extent in $x$. 
In particular, the lowest lying wave functions of the compact harmonic oscillator have very similar support as the non-compact version, as long as the wave function is not spread through the full compact space $-\pi < x < \pi$. This is no longer true when $g$ is sufficiently large. However, since we choose a different value for $x_\text{max}$ when this occurs, the statement is still valid.  
Thus, the analytical expression for the optimal value of $x_\text{max}$ in the regular QHO should hold in the compact QHO, as long as the coupling $g$ is sufficiently small to ensure narrow enough lowest-lying states. Therefore, we define the optimal $x_\text{max}$ value in the non-compact theory to be
\begin{align}
\label{eq:xmaxanalyticalNC}
x^{\rm NC}_{\rm max} =g \ell\sqrt{\frac{\beta_p}{\beta_x}} \sqrt{\frac{2\pi}{2\ell+1}}
\,.
\end{align}
This expression is modified for the compact theory, due to the finite maximum range of the magnetic field. In this case, $x_{\rm max}$ is given by
\begin{align}
\label{eq:xmaxanalyticalC}
x^{\rm C}_{\rm max} = \text{min}\left[x^{\rm NC}_{\rm max},\frac{2\pi \ell}{2\ell+1}\right]
\,,
\end{align}
where the second expression is the $x_{\rm max}$ value in the formulations with $H_P = H^{(1)}_P$ 

\subsection{Numerical Crosschecks}
\begin{figure}[!h]
    \centering
    \includegraphics[width=0.9\textwidth]{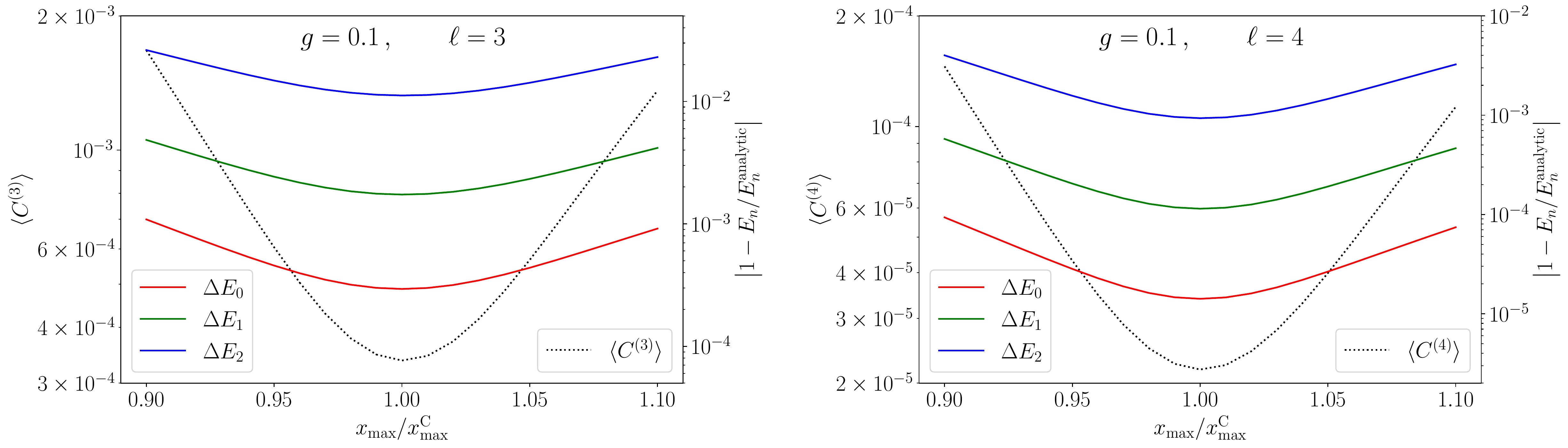}
    \caption{Comparison of $x_\text{max}$ to $x_\text{max}^{\rm C}$ for $\ell = 3, 4$, $g = 0.1$ and $\beta_x, \beta_p =1$. The solid lines show the fractional difference between the energies of the digitizied compact theory and the analytic result; the dashed line shows the deviation from the canonical commutation relation, as defined in Eq.~\eqref{eq:canoncialComm}.}
    \label{fig:ToyModelCompareSmall}
  \end{figure}
 \begin{figure}[!h]
    \centering
    \includegraphics[width=0.9\textwidth]{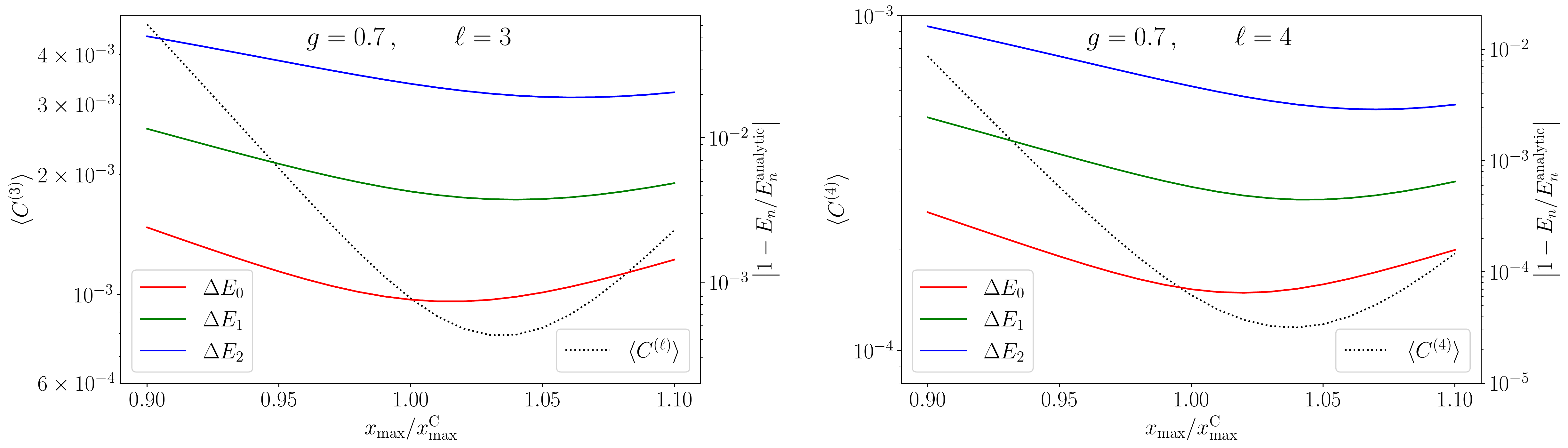}
    \caption{Same as Fig.~\ref{fig:ToyModelCompareSmall}, but with $g = 0.7$. Note that for $\ell = 4$, the value of $x_\text{max}^{\rm C}$ freezes out at $g \sim 0.83$ and the theory becomes equivalent to the KS formulation.}
    \label{fig:ToyModelCompareBig}
  \end{figure}
In order to check the validity of using the analytic expression for $x_\text{max}$, we compare the lowest-lying eigenvalues for the digitized theory to the exact results found in \eqref{app:AnalySol}. Additionally, we calculate the deviation from the canonical commutation relations via the function 
\begin{align}
\label{eq:canoncialComm}
 \left\langle C^{(\ell)} \right\rangle \equiv 1+ i \bra{\Psi_0} [X, P] \ket{\Psi_0}
    \,.
\end{align}
For both comparisons, we vary $b_\text{\rm max}$ in relation to $b_\text{max}^\text{NC}$ in order to test how well the analytical expression for $b_{max}$ performs. This comparison is shown in Fig.~\ref{fig:ToyModelCompareSmall} and Fig.~\ref{fig:ToyModelCompareBig}. For simplicity's sake, we choose $\beta_x = \beta_p = 1$; however, the goodness of our results is independent of $\beta_x, \beta_p$.

Fig.~\ref{fig:ToyModelCompareSmall} shows that for small values of the coupling the minimum of the commutator relation as well as the minima in the fractional energy differences occur essentially at the same point as the analytical result derived in the non-compact theory, indicated by the ratio being 1 on the x-axis. 
While this results is expected, since at such small values of the coupling the compact and non-compact theory should be very similar, the numerical confirmation is still good to see.
At larger values of the coupling, shown in Fig.~\ref{fig:ToyModelCompareBig}, this is no longer true exactly, but the deviations from the analytical result of $x_{\rm max}$ are rather small, and the difference in the fractional energy differences to the optimal result are quite small as well.
Note also that the optimal result depends on the excitation one looks at, which is of course not something that can be handled in a general simulation.

\section{Analytical solution for the non-compact U(1) theory}
\label{app:noncompactU1}
The studies of the previous section demonstrate that, at least for small coupling, the non-compact version of the theory offers important insight on the compact theory. 
This is a useful observation, as the small coupling limit corresponds to the continuum limit of the theory. 
Therefore, it seems prudent to attempt to analytically solve the undigitized version of the non-compact theory. 
In this Section, we take the full $(2+1)$-dimensional pure \uone theory but take the smallest non-trivial system -- four lattice points and periodic boundary conditions, as shown in Fig~\ref{fig:RotorPic}. 
In this case, the Hamiltonian is given by
\begin{align}
H = \underbrace{\frac{2 g^2}{a}\left[R_1^2 + R_2^2+ R_3^2 - R_1 (R_2+R_3)\right]}_{H_R} + \underbrace{\frac{1}{2g^2 a}\left[ B^2_1+ B^2_2 + B^2_3+ (B_1+B_2+B_3)^2\right]}_{H^{NC}_B}
\,,
\end{align}
where all the $B$ variables are related to the plaquette operator via \eqref{eq:MagOpeDef}. 
Noting the exchange symmetry of $R_2 \leftrightarrow R_3$, we implement a change of basis such that the rotor Hamiltonian has no mixed terms. Since this is a unitary transformation, the canonical commutation relations are unchanged. With this change of basis, we have that 
\begin{align}
H_R &= g^2 \left(2 R_1^2+\left(2+\sqrt{2}\right) R_2^2+\left(2-\sqrt{2}\right) R_3^2\right) \nonumber \\
H_B^\text{NC}&=\frac{1}{2g^2}\left(B_1^2+\frac{5-2 \sqrt{2}}{2}B_2^2+ B_2 B_3+\frac{5+2 \sqrt{2}}{2} B_3^2\right)
\,,
\end{align}
where as long as the operator transformations do not alter the canonical commutation relations, we do not distinguish between the old and new operators. 
This Hamiltonian can be further simplified by recalling that the canonical commutation relations are also preserved if the operators are rescaled as
\begin{align}
B_i \rightarrow \alpha_i B_i \qquad R_i \rightarrow \frac{1}{\alpha_i}R_i
\,.
\end{align}
Choosing specific values for $\alpha_i$ transforms the Hamiltonian into
\begin{align}
H_R &= \frac{1}{2}\left(R_1^2 + R_2^2 + R_3^3\right) \nonumber\\
H_B^\text{NC} &=\frac{1}{2}\left(4 B_1^2 + \left(6+ \sqrt{2}\right)B_2^2+ 2\sqrt{2} B_2 B_3 +\left(6- \sqrt{2}\right)B_3^2 \right)
\,.
\end{align}
With this form, any unitary transformation leaves $H_R$ unchanged and so we can now diagonalize $H_B^\text{NC}$. Doing so results in the Hamiltonian
\begin{align}
H = \frac{1}{2}\left(\tilde R_1^2 + \tilde R_2^2 + \tilde R_3^3\right) + \frac{1}{2}\left(4 \tilde B_1^2+ 8 \tilde B^2_2+ 4 \tilde B^2_3\right)
\,,
\end{align}
where we now introduced the `tilde' notation to distinguish these operators from the original ones;  $\{\tilde{R}_i, \tilde{B}_i\}$ still obey the canonical commutation relations. This form of the Hamiltonian is three one-dimensional quantum harmonic oscillators and so the total energy is given by
\begin{align}
E_{n_1, n_2, n_3} &= 2 \left(n_1 + \frac{1}{2}\right) + 2 \sqrt{2} \left(n_2 + \frac{1}{2}\right)+ 2 \left(n_3 + \frac{1}{2}\right)
\,.
\label{eq:NCAnalyticEnergy}
\end{align}
This expression for the energy of the undigitized theory was used in Fig.~\ref{fig:bOptCheck}. It is possible to find the correspond eigenstates in the original basis, making use of the fact that in the basis of eigenvectors of $\tilde{B}_i$, the eigenstates are
\begin{align}
\psi(\tilde b_1, \tilde b_2, \tilde b_3) = N \left(e^{-\tilde b_1^2}H_{n_1}\left[\sqrt{2}\tilde b_1\right]\right)\left(e^{-\sqrt{2}\tilde b_2^2}H_{n_2}\left[\left(2\sqrt{2}\right)^{1/2}\tilde b_2\right]\right)\left(e^{-\tilde b_3^2}H_{n_3}\left[\sqrt{2}\tilde b_3\right]\right)
\end{align}
and the transformation from the `tilde' operators to the original ones are given by
\begin{align}
\tilde b_1 \rightarrow \frac{b_3 - b_2}{2\sqrt{2}g}\, ,\quad \tilde b_2 \rightarrow \frac{b_3 + b_2}{2\sqrt{2}g}\,,  \quad  \tilde b_2 \rightarrow \frac{ 2 b_1+ b_2 + b_3}{2\sqrt{2}g}\, .
\end{align}

\end{document}